\documentclass[%
 reprint,
 superscriptaddress,
 amsmath,amssymb,
 aps,longbibliography
]{revtex4-1}

\usepackage{graphicx}
\usepackage{dcolumn}
\usepackage{bm}
\usepackage{hyperref}
\usepackage[mathlines]{lineno}
\usepackage[bottom]{footmisc}
\usepackage[super]{nth}%

\begin{document}
\raggedbottom

\title{Compact localized states of open scattering media}

\author{Fabrizio Sgrignuoli}
\affiliation{Department of Electrical and Computer Engineering, Boston University, Boston, Massachusetts, 02215, USA.}
\affiliation{Authors with equal contribution}
\author{Malte R{\"o}ntgen}
\affiliation{Zentrum f{\"u}r optische Quantentechnologien, Universit{\"a}t Hamburg, Luruper Chaussee 149, 22761 Hamburg, Germany}
\affiliation{Authors with equal contribution}
\author{Christian V. Morfonios}
\affiliation{Zentrum f{\"u}r optische Quantentechnologien, Universit{\"a}t Hamburg, Luruper Chaussee 149, 22761 Hamburg, Germany}
\author{Peter Schmelcher}
\affiliation{Zentrum f{\"u}r optische Quantentechnologien, Universit{\"a}t Hamburg, Luruper Chaussee 149, 22761 Hamburg, Germany}
\affiliation{The Hamburg Centre for Ultrafast Imaging, Universit{\"a}t Hamburg, Luruper Chaussee 149, 22761 Hamburg, Germany}
\author{Luca Dal Negro}
\email{dalnegro@bu.edu}
\affiliation{Department of Electrical and Computer Engineering, Boston University, Boston, Massachusetts, 02215, USA.}
\affiliation{Division of Material Science and Engineering, Boston University, Boston, Massachusetts, 02215, USA.}
\affiliation{Department of Physics, Boston University, Boston, Massachusetts, 02215, USA}


\begin{abstract}
We study the compact localized scattering resonances of periodic and aperiodic chains of dipolar nanoparticles by combining the powerful Equitable Partition Theorem
(EPT) of graph theory with the spectral dyadic Green's matrix formalism for the engineering of embedded quasi-modes in non-Hermitian open scattering systems in three spatial dimensions. We provide analytical and numerical design of the spectral properties of compact localized states in electromagnetically coupled chains and establish a connection with the distinctive behavior of Bound States in the Continuum. Our results extend the concept of compact localization to the scattering resonances of open systems with arbitrary aperiodic order beyond tight-binding models, and are relevant for the efficient design of novel photonic and plasmonic metamaterial architectures for
enhanced light-matter interaction.
\end{abstract}

\maketitle
The engineering of spatially localized wave excitations in complex scattering media is a fundamental problem that arises in many different contexts of optics and quantum mechanics. In particular, careful design of the symmetry properties of scattering potentials enables the creation of Compactly Localized States (CLS), which are resonant electronic or photonic modes spatially localized over a compact subset of lattice sites \cite{Flach,Maimaiti,Leykam,Leykam2018AP370901PerspectivePhotonicFlatbands,DerzhkoStronglycorrelatedflatband2015,MaltePRB,RealFlatbandlightdynamics2017}. These peculiar modes have received significant attention due to their importance for the design of dispersion-less flat bands (FB) in the spectrum of periodic lattices within the tight-binding model \cite{Flach,Maimaiti,Leykam,Leykam2018AP370901PerspectivePhotonicFlatbands,DerzhkoStronglycorrelatedflatband2015,MaltePRB,RealFlatbandlightdynamics2017}. Moreover, symmetry arguments also play a key role in the manipulation of embedded eigenstates, which are eigenmodes of open large-scale structures with infinite lifetime and diverging quality factors appearing in the radiation continuum  \cite{Neumann,Stillinger,Capasso,Marinica,Hsu,HsuReview,Weimann,Plotnik,Rivera,Doeleman}. In fact, the symmetries of periodic structures, traditionally described by their space groups properties, determine the behavior of such peculiar eigenstates, also referred to as Bound States in the Continuum (BIC) \cite{Neumann,Stillinger,Capasso,Marinica,Hsu,HsuReview,Weimann,Plotnik,Rivera,Doeleman}. Symmetry-induced photonic eigenstates have recently attracted a significant attention in the nano-optics and metamaterials communities due to the many potential applications to ultra-compact and efficient solid-state lasers \cite{Kodigala}, optical sensors \cite{Yanik}, and narrowband filters \cite{Foley}. However, the predictive design of photonic CLS and BICs is currently limited to either the approximate tight-binding approximation or the use of  intensive full-vector numerical simulations of large-scale photonic structures \cite{Leykam,Leykam2018AP370901PerspectivePhotonicFlatbands}.
	
In this letter, we propose a more general framework that captures in three-spatial dimensions the effects of local symmetries in the electromagnetic response of large chains of dipolar scattering nanoparticles with arbitrary geometry, enabling the efficient design of CLS and embedded eigenstates in photonic systems.
Such local symmetries have been treated within a theory of non-local currents and were recently shown to induce a class of CLS in tight-binding networks \cite{MaltePRB,Kalozoumis2014PRL11350403InvariantsBrokenDiscreteSymmetries,Rontgen2017AP380135NonlocalCurrentsStructureEigenstates,Morfonios2017AP385623NonlocalDiscreteContinuityInvariant,Kalozoumis2015PRB9214303InvariantCurrentsLossyAcoustic}. The proposed approach is based on the combination of the rigorous Green's matrix multiple scattering technique and recently established theorems in graph theory here employed to obtain a block partitioning of the dyadic Green's matrix induced by the symmetry properties of a given system under local site permutations. The diagonalization of the Green's matrix is reduced to finding the eigenspectra of smaller matrices, with eigenvectors naturally divided into compact localized and extended states. While valuable as a computational tool for arbitrary symmetric discrete open-scattering media, the proposed approach provides a unified, intuitive, and efficient method for the design of the energy spectra of CLS in both periodic and non-periodic arrays of nanostructures.

Our work applies the powerful Equitable Partition Theorem (EPT) of graph theory to the analysis of the dyadic Green's matrix formalism for the spectral engineering of embedded quasi-modes in non-Hermitian open scattering systems in three spatial dimensions. Moreover, by analyzing the spectral properties of the dyadic Green's matrix we discover a fundamental similarity between CLS quasi-modes (vanishing waves outside a finite subset of the system due to destructive interference \cite{Flach,Maimaiti,MaltePRB,Leykam,Leykam2018AP370901PerspectivePhotonicFlatbands}) and the distinctive behavior of photonic BICs, which is achieved in the limit of photonic systems of infinite optical density. 
Indeed, for infinite optical density, the analyzed structures support distinctive resonances with diverging normalized quality factors $Q/Q_0$ when bound states are gradually decoupled from the continuum. Remarkably, we demonstrate that the vertical mirror symmetry (y-axis mirror) of our system is responsible for the formation of embedded eigenstates regardless of the translational invariance (periodicity) of the system. Indeed, we show that exactly the same $Q/Q_0$ behavior can be obtained by randomly modulating the inter-particle positions along the horizontal x-axis by introducing a white noise structural perturbation in the system.  
Our work demonstrates that the EPT can be successfully applied to the Green's matrix formalism providing a novel and powerful theoretical tool for the analytical investigation of the spectra of embedded quasi-modes based only on local symmetry arguments, beyond the conventional tight-binding approach \cite{MaltePRB}. Our findings are relevant for the efficient design of novel photonic and metamaterials architectures that support scattering resonances with engineered compact localization and BIC behavior leading to enhanced light-matter interactions.
	
The Green's matrix method is a major theoretical tool that accounts for all the scattering orders, such that multiple scattering processes are treated exactly as long as the particles are much smaller than the wavelength. In this limit, each scatterer is characterized by a Breit-Wigner resonance at frequency $\omega_0$ and a resonant width $\Gamma_0$. The quasi-modes of arbitrary open-scattering systems are provided by the eigenvectors of the Green's matrix $\overleftrightarrow{G}$ which, for $N$ vector dipoles, is a $3N\times{3N}$ matrix with components \cite{SkipetrovPRL}:
\begin{equation}\label{Green}
G_{ij}=i\left(\delta_{ij}+\tilde{G}_{ij}\right)
\end{equation}
$\tilde{G}_{ij}$ has the form:
\begin{eqnarray}\label{GreenOur}
\begin{aligned}
\tilde{G}_{ij}=&\frac{3}{2}\left(1-\delta_{ij}\right)\frac{e^{ik_0r_{ij}}}{ik_0r_{ij}}\Biggl\{\Bigl[\mathbf{U}-\hat{\mathbf{r}}_{ij}\hat{\mathbf{r}}_{ij}\Bigr]\\
&- \Bigl(\mathbf{U}-3\hat{\mathbf{r}}_{ij}\hat{\mathbf{r}}_{ij}\Bigr)\left[\frac{1}{(k_0r_{ij})^2}+\frac{1}{ik_0r_{ij}}\right]\Biggr\}
\end{aligned}
\end{eqnarray}
\begin{figure}[t!]
\centering
\includegraphics[width=\linewidth]{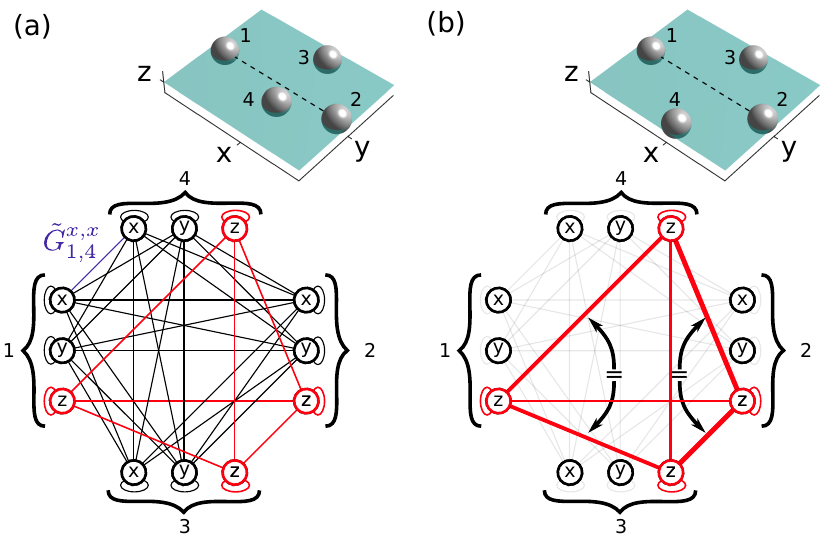}
\caption{Graph decomposition of the Green's matrix for two different configurations shown in the upper part (dotted line denotes the $x$-axis). Each line represents a non-vanishing element of the symmetric matrix (\ref{GreenOur}). Panel (a) shows a generic two-dimensional setup in the $x$-$y$ plane for which $\tilde{G}$ can always be divided into two parts containing, respectively, the $x$-$y$ (black lines) and only the $z$ coupling terms (red lines). Panel (b) displays the effect of a mirror symmetry whit respect to the $x$-axis, inducing a local symmetry in $\tilde{G}$ as  indicated by the thick arrows and red lines.}
\label{Fig1}
\end{figure}
when $i\neq j$ and $0$ for $i=j$. $k_0$ is the wavevector of light, the integer indexes $i, j \in (1,\cdots,N$) refer to different particles, $\textbf{U}$ is the 3$\times$3 identity matrix, $\hat{\mathbf{r}}_{ij}$ is the unit vector position from the $i$-th and $j$-th scatterer while $r_{ij}$ identifies its magnitude.  This formalism has been extensively used for the study of wave transport and localization phenomena in open multiple scattering media \cite{SkipetrovPRL,RusekPRA,Lagendijk,SgrinuoliVogel,WangPRB,WangOL}. Because the matrix (\ref{Green}) is non-Hermitian, its eigenvalues $\Lambda_n$ ($n \in 1,2,\cdots, 3N$) are complex and they completely describe the scattering resonances of the system under study \cite{RusekPRA}. Specifically, the real and the imaginary parts of $\Lambda_n$ are related to the normalized scattering frequency and normalized decay rates, respectively \cite{RusekPRA,SkipetrovPRL}:
\begin{eqnarray}\label{1}
&\Re[\Lambda_n]=(\omega_0-\omega_n)/\Gamma_0\\
&\Im[\Lambda_n]=\Gamma_n/\Gamma_0\label{2}
\end{eqnarray}
The Green's matrix formalism is particularly suitable to study embedded eigenstates whenever each scattering particle is coupled to every other particle due to long-range electromagnetic interactions. Therefore, it is suitable to extend the traditional concept of embedded states \cite{Neumann,Stillinger} to the multiple scattering regime. While the method is limited to vector scattering dipoles, it very-well captures the fundamental multiple scattering physics of large-scale coupled systems without the prohibitive costs associated to traditional numerical techniques that are typically employed for the design of similar electromagnetic structures \cite{Marinica,Hsu,Weimann,Yang,HsuReview}.
	
Originally developed in graph theory \cite{BarrettEquitabledecompositionsgraphs2017}, the EPT exploits a permutation symmetry $\phi$ of a square matrix $M$ to decompose it into smaller matrices $B_{i}$, the eigenvalues of which collectively give those of $M$. The corresponding eigenvectors of $M$ can likewise be constructed from the eigenvectors of $B_{i}$ and can be shown to share the symmetries of $M$. Interestingly, if $\sigma \colon \{1,\ldots{},k\} \to \{ 1,\ldots{},k \}$ acts as the identity operator on a subset $\mathcal{S} \subset \{1,\ldots{}, k\}$, then $M\in \mathbb{C}^{k\times{}k}$ is guaranteed to host at least one CLS whose amplitudes vanishes on $\mathcal{S}$ \cite{Klein2015MCMCC74247LocalSymmetriesMolecularGraphs}. We now apply the EPT to planar configurations that are symmetric with respect to the x-axis (see Fig.\ref{Fig1} (b)). For these configurations, it can be shown that
\begin{align} \label{eq:GSimplification}
\tilde{G}_{ij}^{a,z} &= \tilde{G}_{ij}^{z,a} = 0 \; \forall\; i,j;\; a \in \{x,y\} \\
\tilde{G}_{i j}^{z,z} &= \tilde{G}_{j i}^{z,z} = \tilde{G}_{i \bar{j}}^{z,z} = \tilde{G}_{\bar{j} i}^{z,z} \; \forall \; i \in \mathcal{S}, j\notin \mathcal{S} \nonumber
\end{align}
where $\bar{j}$ denotes the symmetry-mapped counterpart of $j$ and $\mathcal{S}$ identifies the scatterers which are positioned on the symmetry axis.
(\ref{eq:GSimplification}) is equivalent to a local permutation symmetry $\sigma$ of $\tilde{G}$ which pairwise permutes the matrix elements related to the $z$-components of scatterers $j$ with that of $\bar{j}$ and acts as the identity on the $x,y$ components. This allows to apply the EPT, which predicts that $M$ hosts $C = (N -|\mathcal{S}|)/2$ CLS, where $|\mathcal{S}|$ denotes the number of scatterers on the symmetry axis.
The corresponding eigenvalues $\{\tilde{\Lambda}_{CLS}\}$ are obtained by diagonalizing a small matrix $B_{1} \in \mathbb{C}^{C \times C}$. For $C=1$
\begin{equation}\label{BIC}
\tilde{\Lambda}_{CLS} = \tilde{G}_{j j}^{z,z} - \tilde{G}_{j \bar{j}}^{z,z} = 
\left[i-\frac{3}{2}\frac{e^{i k_{0} R}}{k_{0} R}\left(1-\frac{1}{i k_{0} R}-\frac{1}{(k_{0} R)^2}\right)\right]
\end{equation}
with $j$ being either of the two scatterers not lying on the symmetry axis and $R = r_{j \bar{j}}$ being the distance to its mirrored counterpart. In terms of the optical density $\rho \lambda$ (keeping $R = 2/\rho$), where $\rho$ is the linear density across the $x$-axis and $\lambda$ is the optical wavelength, the normalized decay rate $\Gamma_{CLS}/\Gamma_{0} = \Im[\Lambda_{CLS}] \to 0$ for $\rho \lambda \to \infty$, corresponding to the case of embedded bound states in the scattering continuum.
	
\begin{figure}[b!]
\centering
\includegraphics[width=\linewidth]{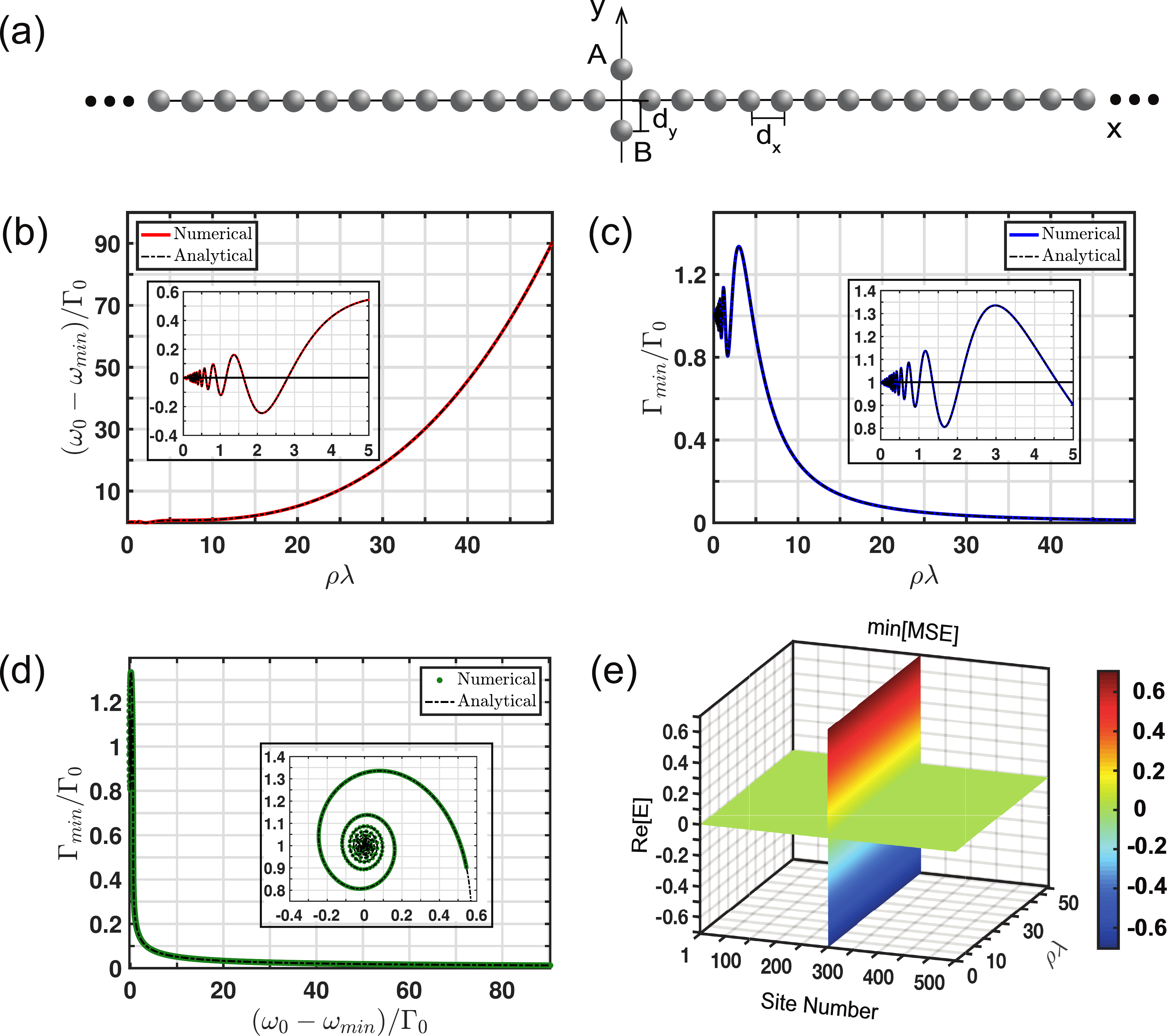}
\caption{The simple benchmark structure, composed by 502 vector electric dipoles, is shown in panel (a). Panels (b-c) show the detuned normalized frequency and the normalized decay rates as function of $\rho\lambda$ corresponding to the Green's matrix eigenvalue with the lowest MSE value (named $min$), respectively. Panel (d) shows how the $min$ complex eigenvalue organizes itself for each considered optical density value. Insets: enlarged view of these trends for low optical density values. The black horizontal lines in the inset of Fig. \ref{Fig2} (b-c) identify the $\omega_{min}$=$\omega_0$ and $\Gamma_{min}$=$\Gamma_0$ point, respectively. Panel (e) displays the real part of the quasi-mode corresponding to the $min$ eigenvalue as a function of the scatterer site number and $\rho\lambda$.}
\label{Fig2}
\end{figure}

In the following, we apply our theory to the simplest possible structure that supports compact localization due to $y$-mirror symmetry. Indeed, it is also very well-known that symmetry-protected BICs can be formed in a system with reflection or rotational symmetry and that their coupling with the extended states is forbidden as long as the symmetry is preserved \cite{HsuReview}. Our test geometry, sketched in Fig. \ref{Fig2} (a), closely resembles the one analyzed in Ref.\cite{Plotnik} and it is characterized by a periodic array of 500 scattering nanoparticles distributed along the x-direction with two additional scatterers located symmetrically above and below the array. Although we present results on a system composed by N=502 scatterers, our findings are independent of N provided that the symmetry along the y-axis is preserved.
Fig. \ref{Fig2} highlights the main concept behind this work, $i.e.$ the possibility to analytically predict the spectrum of CLS quasi-modes by using the EPT on the dyadic Green's matrix (\ref{Green}). In order to immediately identify the CLS resonance in the complex scattering plane of the Green's matrix we computed the Mode Spatial Extent (MSE) parameter for all the modes in the spectrum. This parameter characterizes the spatial extent of a photonic mode \cite{SgrignuoliACS}. Panels (b-d) show the CLS spectrum from Eq. (\ref{BIC} ) (see black-dotted lines) obtained via the EPT Green's matrix decomposition. We also show the CLS spectrum obtained by numerical diagonalization of the Green's matrix, perfectly matching the analytical result. Moreover, panels (b-c) show the detuned normalized frequency and the normalized decay rates corresponding to the Green's matrix eigenvalue with the lowest MSE (referred to as ``the $min$ eigenvalue") as function of the parameter $\rho\lambda$, which corresponds to the optical density of the system (i.e., the number of scattering particles per unit wavelength). The insets display an enlarged view of these trends that feature a fast oscillatory behavior for small values of the optical density ($\rho\lambda$<2). These oscillations of the $min$ eigenvalue around the Breit-Wigner resonance are typical of proximity resonances. Indeed, exactly the same behavior can be observed in a system composed of only two scatterers separated by a distance $d$ \cite{RusekPRA}. The complex scattering plane associated to two electric point dipoles for different distances $d$ is also characterized by two spiral arms associated to the excitation of p-wave and s-wave scattering resonances, respectively. In the limit $d\rightarrow\infty$ they meet at the isolated point characterized by $\omega$=$\omega_0$, $\Gamma$=$\Gamma_0$ \cite{RusekPRA}. To completely clarify the nature of these oscillations we report in panel (d) the normalized decay rate of the $min$ eigenvalue as a function of its normalized detuned frequency. The typical spiral features of proximity resonances are clearly visible in the inset of Fig. \ref{Fig2}(d), demonstrating that in the limit of very low optical density CLS quasi-modes appear as isolated single scattering resonances in the Green's matrix spectrum. In order to confirm that the $min$ eigenvalue corresponds to a CLS quasi-mode, we have evaluated the spatial distribution of its corresponding eigenvector as a function of $\rho\lambda$. Its real part is reported in panel (e). As expected, this eigenvector is non-zero only on the sites of the chain corresponding to the position of the scatterers A and B, as shown in Fig. \ref{Fig2}. This mode is therefore a CLS supported only on the two particles at the center of the chain, with reference to Fig. \ref{Fig2} (a). Our analysis demonstrates that the eigenvalue of the Green's matrix with the lowest MSE  corresponds to a CLS quasi-mode.
\begin{figure}[t!]
\centering
\includegraphics[width=\linewidth]{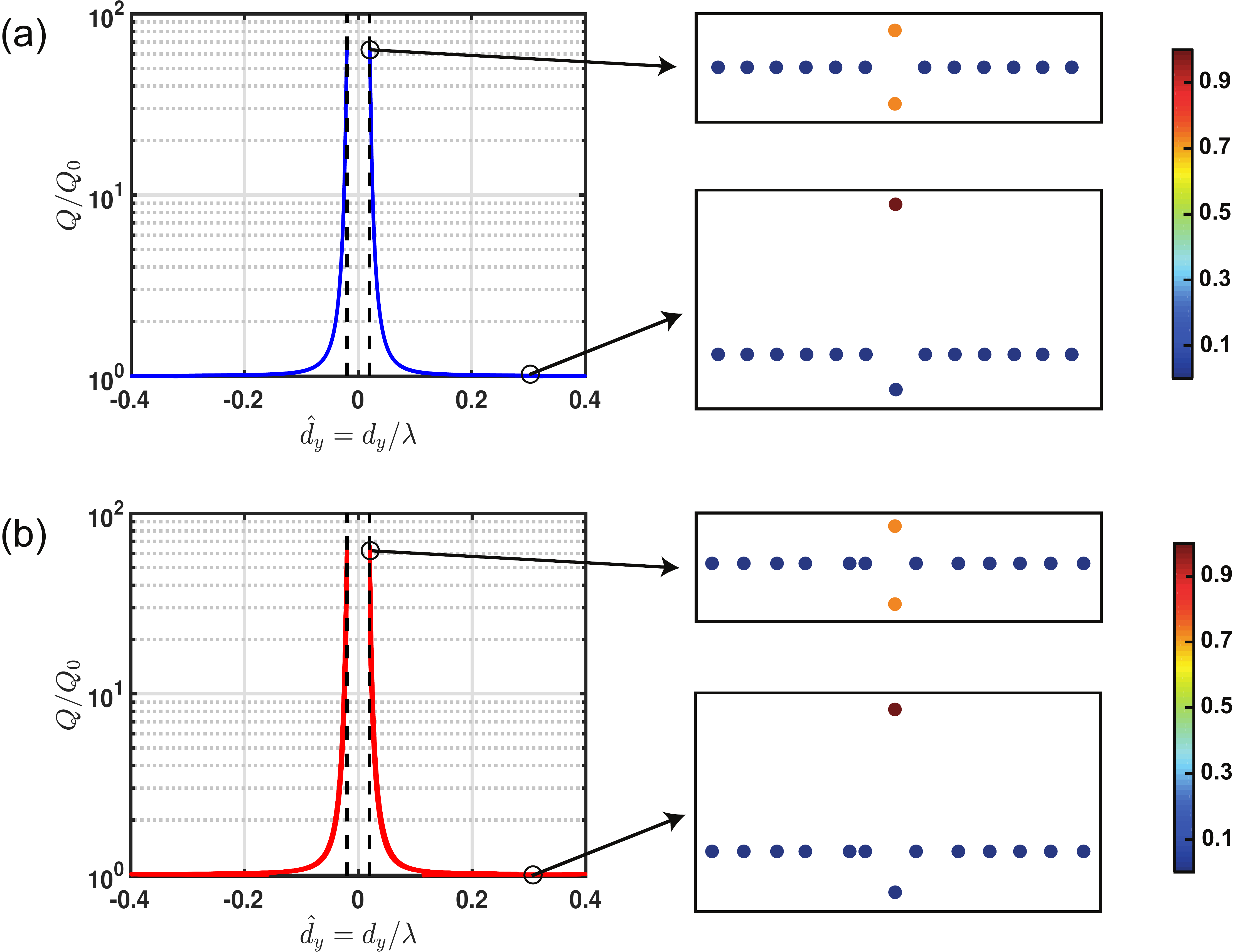}
\caption{Panel (a-b) display the normalized $Q$-factor as a function of position of the scatterers A and B normalized with respect to the optical wavelength, for the periodic and disorder configuration, respectively. These data are obtained by fixing $\rho\lambda$=50. The trend of panel (b) is characterized by 50 different disorder realizations. Representative spatial distribution of the Green's matrix eigenvector, corresponding to the lowest MSE value, are shown for different $\hat{d}_y$ values for both configurations.}
\label{Fig3}
\end{figure}
	
In order to investigate the link between CL quasi-modes and BIC states, we have also analyzed the behavior of the normalized quality factor $Q/Q_0=1/\Im[\Lambda_n]$ of the $min$ eigenvalue as a function of the vertical relative position of the scatterers A and B normalized with respect to the optical wavelength ($\hat{d}_y$). The results of this analysis are reported in Fig. \ref{Fig3} (a-b) for a periodic and a randomly perturbed alignment of scatterers on the $x$-axis, respectively. The results of Fig. \ref{Fig3} are obtained by fixing the optical density equal to 50 and they demonstrate a rapid decrease of the leakage radiation of the minimally extended mode when the $y$-mirror symmetry of the system is progressively restored. This symmetric condition is identified by the black-dotted lines of Fig. \ref{Fig3} (a-b) while the positive and negative values of $\hat{d}_y$ refer to the position of the scatterers A and B, respectively.
Specifically, the quantity $Q/Q_0$ shows a monotonically increasing behavior when the state is gradually decoupled from the continuum, and it diverges in the limit of infinite optical density similarly to what is reported in Ref. \cite{Hsu} for BICs in a photonic crystal slab. In order to demonstrate that this behavior is produced only from the $y$-mirror symmetry, we perturbed the particle positions along the $x$-axis by adding white-noise positional fluctuations\cite{Lord}.
The inter-particle separation fluctuates around the unperturbed distance with relative amplitude $0.5$.
Interestingly, we obtained exactly the same $Q/Q_0$ behavior also in the disorder configuration (as reported in Fig. \ref{Fig3}(b) for $50$ different realizations of the disordered chains), thus demonstrating that the $y$-mirror symmetry is solely responsible for the formation of the CLS states with distinctive BIC behavior. Representative spatial distributions of the Green's matrix eigenvectors corresponding to the lowest MSE values are also reported on the right side of Fig. \ref{Fig3}(a-b). The typical CLS spatial profile is supported only by the central two sites only when the mirror symmetry condition is reached in both configurations. Progressively breaking the $y$-mirror symmetry in our system produces a gradual coupling of the CLS to the continuum states, resulting in radiative losses. 

Our work extends the concept of compact localized states to open electromagnetic scattering systems and fully demonstrates the potential of the EPT graph decomposition theorem applied to the dyadic Green's matrix method for the study of the localization properties and the spectra of symmetric collectively coupled electromagnetic structures. While valuable as a computational framework for arbitrary discrete open-scattering media, the proposed approach provides a unified, intuitive, and efficient method for designing the spectra of CLS and BICs in both periodic and non-periodic arrays of resonant nanostructures.
	
\section*{Funding Information}
Army Research Laboratory (ARL) through the Collaborative Research Alliance (CRA) for MultiScale Multidisciplinary Modeling of Electronic Materials (MSME) under Cooperative Agreement (W911NF-12-2-0023). L.D.N also acknowledges the partial support from the NSF-ECCS EAGER Award 1643118. M.R. gratefully acknowledges financial support by the `Stiftung der deutschen Wirtschaft' in the framework of a scholarship.
\bigskip

%
\end{document}